\documentclass[preprint,aps,pra,showpacs,floatfix]{revtex4}
\usepackage{graphicx}
\usepackage{times}
\usepackage{nicefrac}
\usepackage{amsmath}
\usepackage{amsfonts}
\usepackage{amssymb}
\usepackage{amsthm}
\usepackage{epsf}
\usepackage{bm}
\usepackage{bbm}
\usepackage{times}
\usepackage[english]{babel}
\newcommand{\be}{\begin{eqnarray}}
\newcommand{\ee}{\end{eqnarray}}
\newcommand{\la}{\langle}
\newcommand{\ra}{\rangle}

\newcommand{\eps}{\epsilon}
\newcommand{\veps}{\varepsilon}

\newcommand{\balpha}{\bm{\alpha}}

\newcommand{\bfr}{{\bf r}}

\newcommand{\po}{$2p_{1/2}$}
\newcommand{\pt}{$2p_{3/2}$}
\newcommand{\s}{$2s$}

\newcommand{\kk}{\lambda}

\newcommand{\al}{\alpha}
\newcommand{\az}{\alpha Z}
\newcommand{\aZ}{\alpha Z}

\newcommand{\albi}{\boldsymbol{\alpha}_i}
\newcommand{\albj}{\boldsymbol{\alpha}_j}
\newcommand{\Eres}{E_{\rm res}}
\newcommand{\Ebind}{E_{\rm bind}}
\newcommand{\Eexc}{E_{\rm exc}}

\newcommand{\Vnucl}{V_{\rm nuc}}

\newcommand{\Veff}{V_{\rm eff}}

\newcommand{\xalpha}{x_{\alpha}}
\begin{document}

\title{
QED calculation of~the~$\bm{2p_{1/2}}$-$\bm{2s}$ and~$\bm{2p_{3/2}}$-$\bm{2s}$
transition~energies and~the~ground-state hyperfine~splitting in~lithiumlike~scandium}
%
%
\author{Y.~S.~Kozhedub$^1$, D.~A.~Glazov$^1$, A.~N.~Artemyev$^1$, N.~S.~Oreshkina$^1$,
V.~M.~Shabaev$^1$, I.~I.~Tupitsyn$^1$,  A.~V.~Volotka$^2$, and G.~Plunien$^2$}
\affiliation{
$^1$
Department of Physics, St. Petersburg State University,
Oulianovskaya 1, Petrodvorets, St. Petersburg 198504, Russia \\
$^2$
Institut f\"ur Theoretische Physik, TU Dresden,
Mommsenstra{\ss}e 13, D-01062 Dresden, Germany \\
}
\begin{abstract}
We present the most accurate up-to-date theoretical values of the ${2p_{1/2}}$-${2s}$
and ${2p_{3/2}}$-${2s}$ transition energies and the ground-state hyperfine
splitting in ${\rm Sc}^{18+}$. All two- and three-electron contributions
to the energy values up to the two-photon level are treated in the framework
of bound-state QED without $\aZ$-expansion. The interelectronic interaction
beyond the two-photon level is taken into account by means of the large-scale
configuration-interaction Dirac-Fock-Sturm (CI-DFS) method. The relativistic
recoil correction is calculated with many-electron wave functions in order
to take into account the electron-correlation effect. The accuracy of
the transition energy values is improved by a factor of five compared
to the previous calculations. The CI-DFS calculation of interelectronic-interaction
effects and the evaluation of the QED correction in an effective screening
potential provide significant improvement for the $2s$ hyperfine splitting.
The results obtained are in good agreement with recently published
experimental data.
\end{abstract}
\maketitle
%
%
%
\section{Introduction}
The dielectronic recombination process has proven to be a useful tool
in high-precision measurements of the excitation energy of low-lying levels
in middle-$Z$ lithiumlike systems \cite{madzunkov:PRA:02,kieslich:PRA:04}.
By this method the energy of the ${2p_{3/2}}$-${2s}$ transition in ${\rm Sc}^{18+}$
was determined to be $44.3107(19)$ eV \cite{kieslich:PRA:04}.
A significant improvement of the accuracy was announced recently by M.~Lestinsky
{\it et al}. \cite{les:EGAS,les:HCI}, with the preliminary value of $44.3096(4)$ eV,
and the work on further improvement of this value is in progress \cite{wolf:SPARC}.
In these experiments the energy of the Rydberg resonances $\Eres$ was measured.
The Rydberg state energy $\Ebind$ was evaluated by means of relativistic many-body
perturbation theory (RMBPT). Then the excitation energy of the ion was determined
as $\Eexc=\Eres+\Ebind$. In Ref. \cite{kieslich:PRA:04} the theoretical value
of $\Eexc$ for both ${2p_{1/2}}$-${2s}$ and ${2p_{3/2}}$-${2s}$ transitions
was obtained by means of RMBPT, while for the quantum electrodynamic (QED)
correction the result of Ref. \cite{kim:PRA:91} was taken into account.
The energy resolution achieved in these experiments also allowed for resolving
the $2s$ hyperfine structure. As a result, the $2s$ hyperfine splitting
of lithiumlike scandium was measured to be $6.21(20)$ meV \cite{les:HCI}.

The main goal of the present investigation is to evaluate the ${2p_{1/2}}$-${2s}$
and ${2p_{3/2}}$-${2s}$ transition energies and the ground-state hyperfine splitting
in lithiumlike scandium to the utmost accuracy aiming at a stringent test
of the present state-of-the-art theoretical description of many-electron effects.
Various contributions to the energy of the ${2p}$-${2s}$ transitions are considered
in the next Section. In order to meet the experimental accuracy, rigorous
quantum electrodynamic calculations of the first two orders of perturbation theory
are combined with large-scale configuration-interaction Dirac-Fock-Sturm (CI-DFS)
calculations of the third- and higher-order contributions within the Breit approximation.
The relativistic nuclear recoil corrections are calculated as well. The evaluation
of the hyperfine splitting is accomplished in Section \ref{section:hfs}. The CI-DFS
method is employed to obtain correlation effects of order $1/Z^2$ and higher.
The radiative correction to hyperfine splitting is calculated with an effective
local screening potential.

Relativistic units are used throughout the paper $(\hbar=c=1)$.
%
%
\section{${2p_{1/2}}$-${2s}$ and ${2p_{3/2}}$-${2s}$ transition energies}
\label{section:en}
%
%

We start with the Furry picture, where in the zeroth-order approximation
noninteracting electrons are bound by the Coulomb field of the nucleus.
The Dirac equation yields zeroth-order energies of the one-electron states.
The homogeneously-charged-sphere model of the nucleus is employed with
the value of rms radius $\la r^2 \ra^{1/2}=3.5443(23)$ fm \cite{ADNDT87_185}.

In leading order of the perturbation theory. diagrams of self-energy,
vacuum polarization, and one-photon exchange arise.
Techniques for the evaluation of these corrections nonperturbative in $\az$
have been described in numerous publications (see, e.g., Ref. \cite{mohr:PREP:98}).
For the self-energy correction we interpolate the values presented in Ref. \cite{PRA58_954}
for the $2s$ and $2p_{1/2}$ states and those presented in Ref. \cite{PRA46_4421}
for the $2p_{3/2}$ state. The vacuum-polarization and one-photon exchange
corrections are recalculated in the present work with inclusion of finite-nuclear-size effects.

The second-order contributions can be classified as one-electron two-loop QED
corrections, two-electron QED corrections, and two-photon exchange. Rigorous
calculation of all two-loop QED corrections is a challenging problem. To date,
the dominant part of these corrections was calculated in a wide range of $Z=10-92$
for the $1s$ state only (see Ref. \cite{yerokhin:SESE} and references therein).
Recently, the corresponding results for $2s$, $2p_{1/2}$ and $2p_{3/2}$ states were
presented for high-$Z$ ions \cite{yerokhin:PRL:06}. However, since for low values
of $Z$ the numerical evaluation of the second-order self-energy correction becomes
rather difficult, so far one has to rely on the $\az$ expansion, which reads
\begin{align}
\label{two-loop}
  \Delta E_{\rm two-loop}=m \left( \frac{\al}{\pi} \right)^2 & \frac{(\az)^4}{n^3}
    \Big[
      B_{40} + (\az)B_{50}
\notag\\
      &+ (\az)^2 \left\{ B_{63} L^3 + B_{62} L^2 + B_{61} L + B_{60} \right\} + \cdots
    \Big]
\,,
\end{align}
where $L=\ln[(\az)^{-2}]$. The values of the coefficients for the $2s$ state
can be found in Appendix A of Ref. \cite{RPM77_000001} and for the $2p_{1/2}$
and $2p_{3/2}$ states in Ref. \cite{PRA72_062102}. Since the convergence of
the expansion in $\az$ is known to be rather bad, we assume the uncertainty
to be about $50\%$ in our case.

The two-electron QED corrections are represented by the diagrams of the screened
self-energy and the screened vacuum-polarization. Rigorous evaluation of
the screened self-energy in Li-like ions was performed in Ref. \cite{yerokhin:PRA:99}
for $2s$ and $2p_{1/2}$ states and in Ref. \cite{yerokhin:OS:05} for the $2p_{3/2}$ state.
The screened vacuum-polarization correction was calculated in Ref. \cite{PRA60_45}.
We obtain the corresponding values for $Z=21$ employing the procedure presented
in these works. In order to estimate higher-order (in $1/Z$) terms of the screened
QED correction, the following approximate scheme is used. The first-order QED
correction is evaluated in an effective screening potential and the higher-order
terms are extracted by subtracting the zeroth- and  first-order terms.
The uncertainty of the higher-order screened QED correciton obtained in this way
is assumed to be $100\%$.

The two-photon exchange correction is evaluated within the framework of QED,
following our previous investigations \cite{PRA67_062506,PRA64_032109}.

In order to evaluate the interelectronic-interaction corrections of third
and higher orders we proceed as follows. The Dirac-Coulomb-Breit equation
within the no-pair approximation is solved by means of the large-scale CI-DFS
method \cite{PRA68_022511,PRA72_062503} yielding the many-electron wave functions
and the energy values. The interelectronic-interaction operator employed
in the Dirac-Coulomb-Breit equation reads
\begin{eqnarray}
\label{interaction}
  V_{\rm Breit} = \kk\al \sum_{i>j} \left[ \frac{1}{r_{ij}} - \frac{\albi \cdot \albj}{2r_{ij}}
  - \frac{(\albi \cdot \bfr_{ij}) (\albj \cdot \bfr_{ij})}{2r^{3}_{ij}}\right]
\,,
\end{eqnarray}
where a scaling parameter $\kk$ is introduced in order to separate terms
of different order in $1/Z$ from the numerical results with different $\kk$.
Here $i,j$ enumerate the electrons and $\balpha$ is a vector incorporating
the Dirac matrices. In this way, for small $\kk$, the total energy of the system
can be expanded in powers of $\kk$,
\begin{equation}
  E(\kk)=E_{0}+E_{1}{\kk}+E_{2}{\kk^{2}}+\sum_{k=3}^\infty E_{k} {\kk^k}
\,,
\end{equation}
where
\begin{equation}
\label{derivative}
  E_{k} = \frac{1}{k!} \frac{d^k}{d\kk^k}E(\kk)\Big|_{\kk=0}
\,.
\end{equation}
The higher-order contribution $E_{\geqslant 3}\equiv\sum_{k=3}^{\infty}E_k$
is calculated as $E_{\geqslant 3}=E(\kk=1)-E_{0}-E_{1}-E_{2}$, where the low-order
terms $E_0$, $E_1$, and $E_2$ are determined numerically according to Eq. (\ref{derivative}).
Comparison of $E_{1}$ and $E_{2}$ with corresponding QED results allows us
to conclude that the uncertainty of the higher-order contribution due to the Breit
approximation is less than $0.1\%$.

The full relativistic theory of the nuclear recoil effect can be formulated only
in the framework of QED \cite{shabaev:PRA:98}. To evaluate the recoil effect
within the lowest-order relativistic approximation one can use the operator
(see, e.g., Ref. \cite{shabaev:PRA:98})
\begin{equation}\label{recoil}
  H_M = \frac{1}{2M} \sum_{i,j} \left[ \boldsymbol{p}_i\cdot\boldsymbol{p}_j
    - \frac{\az}{r_i} \left( \albi+\frac{(\albi\cdot\boldsymbol{r}_i)\boldsymbol{r}_i}
    {r^{2}_{i}} \right) \cdot\boldsymbol{p}_j \right],
\end{equation}
where $M$ is the nuclear mass and $\boldsymbol{p}_i$ is the momentum operator
acting on the ith electron. The expectation value of $H_M$ on the many-electron
wave function of the system, obtained by the CI-DFS method, yields the recoil
correction to the energy levels in all orders of $1/Z$ within the $(\az)^4{m^2}/{M}$
approximation. The electron-correlation effects contribute to about $20\%$
of the total value and have to be taken into account in order to achieve
the desirable accuracy. The one- and two-electron recoil corrections of higher
orders in $\az$ are taken from Refs. \cite{artemyev:PRA:95, JPB28_5201}.
The recoil correction of the next order in $m/M$ is negligible in the case
under consideration.

All contributions to the transition energies considered above are collected
in Table I. For comparison, previous theoretical results and available
experimental data are presented as well. As one can see from the table,
the theoretical values of the transition energies reported in this paper
are about five times more precise than those in Ref. \cite{kieslich:PRA:04}
and agree well with the experiments. Further improvement of the theoretical
accuracy can be achieved by more accurate calculations of the higher-order
screened QED effects.
%
%
\section{Hyperfine splitting}
\label{section:hfs}

The ground-state hyperfine splitting of a lithiumlike ion is conveniently written as
\begin{eqnarray}
\label{eq:hfs}
  \Delta E_\mu &=& \frac{1}{6}\,\alpha\,(\aZ)^3\,\frac{m}{m_p}\,
    \frac{\mu}{\mu_N}\,\frac{2I+1}{2I}\,\frac{1}{(1+\frac{m}{M})^3}\,mc^2
\nonumber\\
  && \times
    \left[ A(\aZ)(1-\delta)(1-\veps) +
    \frac{1}{Z}B(\aZ) + \frac{1}{Z^2}C(Z,\aZ)
    + x_{\rm{rad}} \right]
\,,
\end{eqnarray}
where $m_p$ is the mass of the proton, $\mu$ and $I$ are the nuclear magnetic moment
and spin, and $\mu_N$ denotes the nuclear magneton. The one-electron relativistic
factor $A(\aZ)$ can easily be derived from the Dirac equation utilizing virial
relations \cite{shabaev:91}. The finite-nuclear-size correction $\delta$
is evaluated numerically employing the homogeneously-charged-sphere model
for the nuclear-charge distribution. The Bohr-Weisskopf correction $\veps$,
arising due to the nonpointlike nuclear magnetization distribution, is evaluated
within the single-particle nuclear model \cite{shabaev:94,shabaev:pra:97}.

The first-order interelectronic-interaction correction described by the function
$B(\az)$ is evaluated in the rigorous QED approach \cite{shabaeva:95}.
The dual-kinetic-balance (DKB) approach \cite{shabaev:04:prl} is employed
to construct the complete set of one-electron wave functions from
the B splines. The finite distributions of the nuclear charge and the nuclear
magnetization are taken into account. The latter is introduced via the replacement
of $1/r^2$ with $F(r)/r^2$ in the hyperfine interaction matrix elements. The explicit
form of the function $F(r)$ can be found in Refs. \cite{tup:02,zherebtsov:00}.
The higher-order correction $C(Z,\aZ)/Z^2$ is obtained in the framework of
the large-scale CI-DFS method.

The QED correction $x_{\rm{rad}}$ is evaluated in one-loop approximation
with an effecitve non-Coulomb binding potential $\Veff$, which partly takes into
account the interelectronic-interaction effects. It is taken in the following
form \cite{slater,kohn-sham}
\begin{eqnarray}
\label{eq:Vscr-dft}
  \Veff(r) = \Vnucl(r) + {\alpha} \int_0^{\infty}dr' \frac{1}{r_>} \rho(r')
  - \xalpha\,\frac{\alpha}{r} \left( \frac{81}{32\pi^2} r \rho(r) \right)^{1/3}.
\end{eqnarray}
Here $\rho$ is the total electron density, including the $(1s)^2$ shell and
the $2s$ electron. The parameter $\xalpha$ is taken to be $\xalpha=2/3$, which
corresponds to the Kohn-Sham potential. To provide a proper asymptotic behavior,
the potential $\Veff$ should be corrected at large $r$ \cite{latter}.
The one-electron spectrum of the Dirac equation with $\Veff$ is constructed
by means of the DKB method \cite{shabaev:04:prl}. Since the potential $\Veff$
is assumed to be self-consistent, the standard iteration procedure is employed.
The calculations performed are very similar to our recent calculations of
the one-loop QED corrections to the $g$~factor of Li-like ions \cite{glazov:pla:06}.
We mention also that the evaluation of the QED corrections to the hyperfine
structure with an effective screening potential was performed in the past
for the case of lithiumlike bismuth \cite{sapirstein:03}.

The individual contributions to the hyperfine splitting in litiumlike scandium
are listed in Table II. For each contribution the corresponding term
in the square brackets in Eq. (\ref{eq:hfs}) is explicitly written.
For comparison, the experimental value from Ref. \cite{les:HCI} as well as
the previously published results by Shabaev {\it et al.} \cite{shabaev:97}
and by Boucard and Indelicato \cite{boucard:00} are presented. The accuracy
of the present result is twice better than that of Ref. \cite{shabaev:97}
and about two orders of magnitude higher than the experimental one.
%
%
\section{Conclusion}
\label{section:con}

In this paper we have presented {\it ab initio} QED evaluations of the ${2p_{1/2}}$-${2s}$
and ${2p_{3/2}}$-${2s}$ transition energies in lithiumlike scandium, where
the most accurate experimental data for middle-$Z$ lithiumlike ions have been achieved.
All presently available contributions to the transition energies are collected.
Except for the one-electron two-loop correction, all other terms up to the two-photon
level are treated within the framework of bound-state QED to all orders in $\az$.
The third- and higher-order interelectronic-interaction effects are accounted for
within the Breit approximation using large-scale CI-DFS calculations.
The relativistic recoil corrections are evaluated as well. As a result, the total
theoretical accuracy is improved by a factor of 5 compared to the previous
calculations.

The ground-state hyperfine splitting of lithiumlike scandium has been calculated.
The interelectronic-interaction correction to the first order in $1/Z$ is evaluated
within the framework of QED. The higher-order electron-correlation effects are calculated
using the large-scale CI-DFS method. The one-loop radiative corrections are calculated
with an effective screening potential. The theoretical value of the hyperfine splitting
is improved in comparison with the previous results.
%
%
\acknowledgments
This work was supported in part by RFBR (Grant No. 07-02-00126), INTAS-GSI
(Grant No. 06-1000012-8881), GSI, and DFG.
Y.S.K. and N.S.O. acknowledge support by the Dynasty Foundation.
The work of N.S.O. and D.A.G. was supported by DAAD.
Y.S.K. acknowledges the support from GSI and from St. Petersburg Government
(Grant No. M06-2.4D-295).
D.A.G. also acknowledges support from the St. Petersburg Government
(Grant No. M06-2.4K-280).
A.N.A., A.V.V., and G.P. acknowledge financial support from DFG and GSI.
%
%
\begin{table}
\label{tab:en}
\caption{
Individual contributions to the ${2p_{1/2}}$-${2s}$ and ${2p_{3/2}}$-${2s}$
transition energies in Li-like scandium, in eV. For comparison, the theoretical
result from Ref. \cite{kieslich:PRA:04} and the experimental values, obtained
via optical spectroscopy \cite{pl:suckewer80} and via the dielectronic recombination
process \cite{kieslich:PRA:04,les:EGAS}, are presented.
}
\linespread{1}
\begin{center}
\begin{tabular}{lr|r@{}l|r@{}lr@{}l}
\hline
\hline
&&
\multicolumn{2}{c}{\po-\s} &
\multicolumn{2}{c}{\pt-\s}
\\
\hline
Dirac value (extended nucleus) &&
    $-$0.&00237 &
       8.&93553 \\
One-photon exchange &&
      41.&89788 &
      38.&90847 \\
Self-energy &&
    $-$0.&2871(3) &
    $-$0.&2679(3) \\
Vacuum-polarization &&
       0.&01979 &
       0.&01989 \\
Two-photon exchange &&
    $-$3.&5683(2) &
    $-$3.&2388(2) \\
Screened QED &&
       0.&0387(20) &
       0.&0331(20) \\
Three- and more-photon exchange &&
    $-$0.&0594(3) &
    $-$0.&0713(3) \\
Two-loop QED &&
       0.&00011(5) &
       0.&00008(4) \\
Recoil &&
    $-$0.&00991(2) &
    $-$0.&01001(2) \\
\hline
Theory: &this work  &
       38.&0294(21) &
       44.&3091(21) \\
&
S.~Kieslich {\it et al}. \cite{kieslich:PRA:04}&  
       38.&0261(100) &
       44.&3089(100) \\
Experiment: &
S.~Suckewer {\it et al}. \cite{pl:suckewer80} &
       38.&02(4)   &
       44.&312(35) \\
&
S.~Kieslich {\it et al}. \cite{kieslich:PRA:04} &
       &    &
       44.&3107(19) \\
&
M.~Lestinsky {\it et al}. \cite{les:EGAS} &
       &    &
       44.&3096(4) \\
\hline
\hline
\end{tabular}
\end{center}
\end{table}
%
%
\begin{table}
\label{tab:hfs}
\caption{
Individual contributions to the ground-state hyperfine splitting of lithiumlike scandium, in meV.
Comparison with the available theoretical and experimental values in terms of the wavelength
$\lambda$ is presented.
}
\linespread{1}
\begin{center}
\begin{tabular}{lr|r@{}l}
\hline
\hline
Dirac value                                      &$A(\aZ)$                 &   6.&9650     \\
Finite-nuclear-size correction                   &$-\delta A(\aZ)$         &$-$0.&0224(3)  \\
Bohr-Weisskopf correction                        &$-\eps A(\aZ)(1-\delta)$ &$-$0.&0064(32) \\
Interelectronic interaction, $1/Z$               &$B(\aZ)/Z$               &$-$0.&8817     \\
Interelectronic interaction, $1/Z^2$ and higher  &$C(Z,\aZ)/Z^2$           &   0.&0150(2)  \\
QED (with screening)                             &$x_{\rm rad}$            &$-$0.&0061(6)  \\
\hline
Total theory, this work                          &$\Delta E_\mu$           &   6.&0633(33) \\
\hline
Wavelength, this work                            &$\lambda$                &   0.&020448(10) cm \\
Theory: & Shabaev {\it et al.} \cite{shabaev:97}                           &   0.&020450(20) cm \\
        & Boucard and Indelicato \cite{boucard:00}                         &   0.&020403 cm     \\
Experiment: & Lestinsky {\it et al.} \cite{les:HCI}                        &   0.&0200(7) cm \\
\hline
\end{tabular}
\end{center}
\end{table}
%
%
%
%

%
\end{document}